\documentclass[letterpaper,american,twocolumn,showpacs,pre,aps,superscriptaddress]{revtex4}
\usepackage[latin1]{inputenc}
\usepackage{bm}
\usepackage{multirow,amssymb,amsbsy,amsmath}
\usepackage{amsmath}
\usepackage{graphicx}
\usepackage{amssymb}
\usepackage{hyperref}
\makeatletter
\usepackage{pifont}
\makeatother

\begin{document}

\title{Structure of completely positive quantum master equations with memory kernel}

\author{Heinz-Peter Breuer}

\email{breuer@physik.uni-freiburg.de}

\affiliation{Physikalisches Institut, Universit\"at Freiburg,
             Hermann-Herder-Strasse 3, D-79104 Freiburg, Germany}

\author{Bassano Vacchini}

\email{vacchini@mi.infn.it}

\affiliation{Dipartimento di Fisica dell'Università di Milano and
             INFN Sezione di Milano, Via Celoria 16, I-20133 Milano, Italy}

\date{\today}

\begin{abstract}
Semi-Markov processes represent a well known and widely used class
of random processes in classical probability theory. Here, we
develop an extension of this type of non-Markovian dynamics to the
quantum regime. This extension is demonstrated to yield quantum
master equations with memory kernels which allow the formulation
of explicit conditions for the complete positivity of the
corresponding quantum dynamical maps, thus leading to important
insights into the structural characterization of the non-Markovian
quantum dynamics of open systems. Explicit examples are analyzed
in detail.
\end{abstract}


\maketitle

\section{Introduction}\label{sec:introduction}

Dissipation, damping and dephasing phenomena in the dynamics of
open quantum systems can often be modelled through the standard
techniques of the theory of quantum Markov processes in which the
open system's density matrix is governed by a quantum master
equation with Lindblad structure \cite{Gorini1976a,Lindblad1976a}.
However, in the description of complex quantum mechanical systems
one encounters in many physically relevant cases a complicated
non-Markovian behavior \cite{Breuer2007} that cannot be described
by means of these standard methods. In fact, non-Markovian systems
feature strong memory effects, finite revival times caused by
long-range correlation functions, and non-exponential damping and
decoherence which generally render impossible a theoretical
treatment through a dynamical semigroup (see, e.g.,
Refs.~\cite{Budini2004a,Budini2005a,Budini2005b,Budini2006a,Breuer2006a,Breuer2007a,Vacchini2008a,Piilo2008a,Ferraro2008a,Krovi2007a}).
As a consequence the analysis of non-Markovian quantum dynamics is
extremely demanding. Even in the regime of classical probability
theory it is difficult to formulate general equations of motion
for the probability distributions of non-Markovian processes. In
quantum mechanics the situation is even more involved since the
classical condition of the preservation of the positivity for the distribution
functions is to be replaced by the stronger condition of complete
positivity of the resulting quantum dynamical maps.

In classical probability theory and the theory of stochastic
processes there exists however a well established and widely used
class of non-Markovian processes, namely the class of semi-Markov
processes
\cite{Feller1964a,Feller1971,Cox1965,Nollau1980,Hughes1995}. It is
thus natural to investigate possible generalizations of this type
of processes to the quantum case. Recently we have proposed such a
generalization, leading to the concept of a quantum semi-Markov
process \cite{Breuer2008a}. In the present paper we elaborate the
details of this approach and indicate a number of further examples
of applications of the theory. The class of quantum processes
constructed here is demonstrated to yield generalized master
equations with memory kernel, and to allow the mathematical
formulation of necessary and even necessary and sufficient
conditions which ensure the complete positivity of the
corresponding quantum dynamical maps. Indeed the formulation of such
conditions for non-Markovian master equations is a highly nontrivial
task \cite{Barnett2001a,Budini2004a,Maniscalco2007a,Wilkie2009a}.

The paper is organized as follows. Section
\ref{sec:class-semi-mark} contains a short introduction into the
theory of classical semi-Markov processes. We define the
fundamental quantities, such as the semi-Markov matrix, the
survival probabilities and the waiting time distributions, derive
the generalized master equation and the structure of the classical
memory kernel, and discuss several examples for classical memory
functions and waiting time distributions. The generalization of
these concepts to the quantum case is developed in
Sec.~\ref{sec:quantum-semi-markov}. We introduce a class of
quantum master equations with memory kernel and formulate
explicitly the corresponding conditions for the complete
positivity of the quantum time evolution. A number of examples and
applications is also discussed. Finally, some conclusions are
drawn in Sec.~\ref{sec:conclu}.

\section{Classical semi-Markov processes}\label{sec:class-semi-mark}
In the present section we want to give a brief introduction to
classical semi-Markov processes, focusing on the basic quantities
necessary in order to describe and uniquely determine such
processes. In particular building on these quantities we will be
able to explicitly derive a generalized master equation for the
time evolution of the conditional transition probabilities of the
process, which is the starting point for a generalization to the
quantum case. General references to the subject are typically
found in the mathematics literature
\cite{Feller1964a,Feller1971,Cox1965} (see also the monograph
\cite{Nollau1980}), even though examples of classical semi-Markov
processes have been extensively studied in the physics literature
under the name continuous time random walk (see \cite{Hughes1995}
for a comprehensive treatment and references therein).

\subsection{Semi-Markov matrix}
\label{sec:semi-markov-matrix}

Semi-Markov processes naturally generalize Markov processes by
combining the theory of Markov chains and of renewal processes
\cite{Ross2007}. In a Markov chain a system jumps among different
states according to certain probabilities depending on departure
and arrival state, the time spent in a given state being
immaterial. A renewal process is instead a counting process in
which the times among successive events are independent
identically distributed random variables characterized by an
arbitrary common waiting time distribution, the adjective renewal
stressing the fact that the process starts anew at every step. If
this waiting time distribution is of exponential type one obtains
as a special case of renewal process a Poisson process, in fact
the exponential is the only memoryless distribution leading to a
Markov counting process. In this case knowing that a system has
already been in a state for a given amount of time provides no
additional information on the expected time of the next jump.  By
combining the two features a semi-Markov process describes a
system moving among different states according to fixed transition
probabilities, so that the sequence of visited states forms a
Markov chain, spending a random time in each state.  These random
sojourn times however are described by a waiting time distribution
which is not necessarily of exponential type, as in a Markov
process, and which might depend both on the present state and on
the immediately following one. If one only considers the different
states visited by a semi-Markov process one recovers a Markov
chain, while if the state space is reduced to a single point one
recovers a renewal process.

A semi-Markov process is uniquely determined introducing a
so-called semi-Markov matrix $Q_{mn}(\tau)$, which gives the
probabilities for a jump from a state $n$ to a state $m$ in a time
$\tau$. More precisely, given that the process arrived in the
state $n$ at time $t$, $Q_{mn}(\tau)$ denotes the probability that
it jumps to the next state $m$ no later than time $t+\tau$. The
semi-Markov matrix can be expressed through the corresponding
densities $q_{mn}(\tau)$ defined by
\begin{equation} \label{eq:1}
 dQ_{mn}(\tau) = q_{mn}(\tau)d\tau,
\end{equation}
which represent a collection of state dependent waiting time
distributions. If a jump eventually occurs with certainty the
following normalization holds
\begin{equation}
   \label{eq:2}
   \sum_{m}\int_{0}^{+\infty} d\tau\, q_{mn} (\tau)=1.
\end{equation}
In terms of the state dependent waiting time distribution $q_{mn}
(\tau)$ one can naturally introduce the survival probability
\begin{equation}
   \label{eq:4}
   g_{n} (\tau) =1-\sum_{m}\int_0^{\tau}ds\, q_{mn} (s),
\end{equation}
that is the probability not to have left state $n$ by time $\tau$.
For the special case in which the waiting time distribution for
the next jump to take place does only depend on the initial state
one has the factorization
\begin{equation}
   \label{eq:3}
   q_{mn}
(\tau)=\pi_{mn}f_{n} (\tau)
\end{equation}
with $\pi_{mn}$ the transition probabilities of the corresponding
Markov chain satisfying $\sum_{m}\pi_{mn}=1$, and $f_{n}(\tau)$ a
normalized waiting time distribution. Correspondingly one also
has the factorization $Q_{mn} (\tau)=\pi_{mn}F_{n} (\tau)$, with
$F_{n} (\tau)$ the cumulative distribution function providing the
probability of a jump out of state $n$ in a time $\tau$. If the
system can get stuck in some state $n$, the corresponding
$f_{n}(\tau)$ is not normalized to one and \eqref{eq:2} becomes a
strict inequality.

It is of interest to consider the form of $q_{mn} (\tau)$
corresponding to a Markov process. Such a process is recovered for
a factorizing expression of the form
\begin{equation}
   \label{eq:5}
   q_{mn} (\tau)=\pi_{mn} \lambda_{n} e^{-\lambda_{n}\tau},
\end{equation}
with corresponding survival probability given by
\begin{equation}
   \label{eq:6}
   g_{n} (\tau) = e^{-\lambda_{n}\tau}.
\end{equation}
Denoting by $\hat h (u)$ the Laplace transform of a function $h
(\tau)$ defined on the positive real line,
\begin{equation}
   \label{eq:10}
   \hat h (u)=\int_{0}^{+\infty} d\tau\, h (\tau) e^{-u\tau},
\end{equation}
we observe for later use that in Laplace representation
semi-Markov matrix and survival probability for a Markov process
read
\begin{equation}
   \label{eq:12}
   \hat q_{mn} (u)=\pi_{mn} \frac{\lambda_{n}}{u+\lambda_{n}}
\end{equation}
and
\begin{equation}
   \label{eq:13}
   \hat g_{n}=\frac{1}{u+\lambda_{n}},
\end{equation}
respectively. This choice corresponds to an exponential waiting
time distribution $f_{n} (\tau)=\lambda_{n} e^{-\lambda_{n}\tau}$,
leading to the following memoryless property. Let us denote by
$\tau_{n}$ the random variable giving the time spent in state $n$,
and let us consider the conditional probability for a jump out of
$n$ to take place after a time $t+s$, given that no jump has taken
place up to time $s$, one immediately has from~\eqref{eq:6}
\begin{equation} \label{eq:7}
 P\{\tau_{n}>t+s | \tau_{n}>s\} =
 \frac{P\{\tau_{n}>t+s\}}{P\{\tau_{n}>s\}}=e^{-\lambda_{n} t},
\end{equation}
so that this conditional probability does not depend on the time
already spent in site $n$. This lack of memory only holds for an
exponential distribution, whose survival probability is given by
the simple exponential~\eqref{eq:6}. For all other possible
choices of the semi-Markov matrix semi-Markov processes are indeed
non-Markovian.

\subsection{Generalized master equation}\label{sec:gener-mast-equat}
We now want to obtain a generalized master equation for the
conditional transition probabilities of a classical semi-Markov
process starting from the central quantity given by the semi-Markov
matrix $q_{mn} (\tau)$. Such a generalized master equation is the
counterpart for the non-Markovian case of the Pauli master equation,
which is reobtained as a special case when memory effects are absent.
To do this we will follow a straightforward and intuitive path,
exploiting an analog of the Kolmogorov forward equation for standard
Markov processes, written in Laplace representation. Another slightly
more indirect route can be found in \cite{Gillespie1977a}, which
already represented an endeavour to give a simple derivation of the
generalized master equation. The point is not entirely trivial, as can
be seen from the amount of literature devoted in the physics community
to relate continuous time random walks, which provide examples of
semi-Markov processes, to generalized master equations (see, e.~g.,
\cite{Allegrini2003a} and references therein).

As a starting point we consider the Kolmogorov forward equations
for a Markov process, which can be immediately written down using
arguments of probabilistic nature \cite{Feller1971}. We denote by
$T_{mn}(t)$ the conditional transition probability, i.~e. the
probability for the process to be in the state $m$ at time $t$
under the condition that it started in state $n$ at time zero.
These quantities obey the equation
\begin{equation}
   \label{eq:9}
   T_{mn} (t)=\delta_{mn}e^{-\lambda_{n}t}+\int_{0}^{t} d\tau\,
   \sum_{k} e^{-\lambda_{m} (t-\tau)} \pi_{mk}\lambda_{k}T_{kn} (\tau),
\end{equation}
the two terms on the r.h.s corresponding to contributions in which
the system has performed zero or at least one jump respectively.
Thus the first expression on the r.h.s. gives the probability not
to have left state $n$, expressed by means of the survival
probability of a Markov process $g_{n} (\tau) =
e^{-\lambda_{n}\tau}$. The second expression argues on the last
jump performed, summing over paths in which the system goes from
state $n$ to a state $k$ in a time $\tau$ and makes at this point
his last jump from $k$ to $m$, with probability density
$\pi_{mk}\lambda_{k}$, dwelling there for the remaining time
$t-\tau$. This equation is most easily dealt with in Laplace
representation, coming to
\begin{equation} \label{eq:11}
 \hat T_{mn} (u) =
 \delta_{mn}\hat g_{m} (u)+\sum_{k}\hat g_{m} (u)\pi_{mk}\lambda_{k} \hat T_{kn} (u),
\end{equation}
and further recalling~\eqref{eq:12} and~\eqref{eq:13}
\begin{equation}
   \label{eq:14}
\hat T_{mn} (u)=\delta_{mn}\hat g_{m} (u)+\sum_{k}\hat g_{m} (u)
\frac{\hat q_{mk} (u)}{\hat g_{k} (u)} \hat T_{kn} (u),
\end{equation}
where the ratio between Laplace transform of semi-Markov matrix
and survival probability appears, which we will generally denote
as
\begin{equation}
   \label{eq:15}
   \hat W_{mk} (u)=\frac{\hat q_{mk} (u)}{\hat g_{k} (u)}.
\end{equation}
We have thus recast the Kolmogorov forward equations in a form
where only the semi-Markov matrix and the related survival
probability appear, starting from their specific expressions for
the case of a Markov process. We now extend these equations to
allow for a general semi-Markov matrix, thus obtaining a set of
equations playing the role of Kolmogorov forward equations for a
semi-Markov process, first obtained by Feller \cite{Feller1964a}.
Recalling that due to Eq.~\eqref{eq:4} the general expression for
the Laplace transform of the survival probability in terms of the
semi-Markov matrix is given by
\begin{equation}
   \label{eq:16}
   \hat g_{n} (u)=\frac{1-\sum_{m} \hat q_{mn} (u)}{u},
\end{equation}
and subtracting the term $\sum_{k} \hat q_{km} (u)\hat T_{mn} (u)$
from both sides of Eq.~\eqref{eq:14} one comes to
\begin{multline}
   \label{eq:17}
\hat T_{mn} (u)[1-\sum_{k} \hat q_{km} (u)]=\delta_{mn}\hat g_{m} (u)
\\
+\hat g_{m} (u)\sum_{k}
\frac{\hat q_{mk} (u)}{\hat g_{k} (u)} \hat T_{kn} (u)
-\sum_{k} \hat q_{km} (u)\hat T_{mn} (u),
\end{multline}
and finally diving by $\hat g_{m} (u)$ one obtains
\begin{equation}
   \label{eq:18}
u\hat T_{mn} (u)-\delta_{mn}=\sum_{k}
[\hat W_{mk} (u)\hat T_{kn} (u)-\hat W_{km} (u)\hat T_{mn} (u)].
\end{equation}
Taking the inverse Laplace transformation of this equation and
using $T_{mn} (0)=\delta_{mn}$ one is thus immediately led to the
generalized master equation
\begin{multline}
   \label{eq:19}
 \frac{d}{dt} T_{mn} (t) = \int_0^t d\tau \sum_k
 \Big[ W_{mk} (\tau) T_{kn} (t-\tau)
 \\
- W_{km} (\tau)T_{mn} (t-\tau) \Big].
\end{multline}
Denoting by $P_{n}(t)$ the probability to be in state $n$ at time
$t$ starting from a fixed state at the initial time zero one can
also write this equation as
\begin{multline} \label{eq:20}
 \frac{d}{dt} P_n (t) = \int_0^t d\tau \sum_m
 \Big[ W_{nm} (\tau) P_m (t-\tau) \\
 - W_{mn} (\tau)P_n (t-\tau) \Big].
\end{multline}

\subsection{Classical Memory kernel}\label{sec:memory-kernel}
The matrix of functions $W_{nm} (t)$ can be naturally called
classical memory kernel, and is given by the inverse Laplace
transform of Eq.~\eqref{eq:15}, expressed in the time domain
through
\begin{equation}
   \label{eq:20bis}
 q_{mn} (\tau) =\int_0^{\tau} ds\, W_{mn} (s)g_{n}(\tau-s)\equiv (W_{mn} \ast
 g_{n})(\tau),
\end{equation}
where $\ast$ denotes as usual the convolution product, or more
compactly in terms of the Laplace transformed quantities
\begin{equation} \label{eq:26}
 \hat W_{mn} (u) = \frac{\hat q_{mn} (u)}{\hat g_{n} (u)}
 =\frac{u\hat q_{mn} (u)}{1-\sum_{l} \hat q_{ln} (u)}.
\end{equation}
As one immediately checks, in the Markovian case the memory kernel
is given by a matrix of positive constants times a delta function
\begin{equation}
   \label{eq:8}
   W_{mn} (t)=\Gamma_{mn}2\delta (t),
\end{equation}
with
\begin{equation}
   \label{eq:21}
   \Gamma_{mn}=\pi_{mn}\lambda_{n},
\end{equation}
thus satisfying
\begin{equation}
   \label{eq:22}
   \Gamma_{mn}\geq 0, \qquad \sum_{m} \Gamma_{mn} = \lambda_n,
\end{equation}
leading to the usual Pauli master equation
\begin{equation}
   \label{eq:23}
          \frac{d}{dt}
P_n (t)
=   \sum_m
 \left[ \Gamma_{nm}
P_m (t)
 - \Gamma_{mn}P_n (t) \right].
\end{equation}
Note in particular that the positivity and the normalization of
the coefficients $\Gamma_{nm}$ naturally allow an interpretation
as transition probabilities per unit time, i.~e., as transition
rates, a simple picture which is no more available in the general
case. In fact, the functions $W_{nm} (t)$ can take on negative
values even when obtained from a well-defined semi-Markov matrix,
as we will show with simple examples.

To do this let us first consider in detail the situation described
in Eq.~\eqref{eq:3}, corresponding to factorized contributions in
the semi-Markov matrix \cite{Gillespie1977a}. We note that in this
case the survival probability simply reads
\begin{equation}
   \label{eq:24}
   g_{n} (\tau) =1-\int_0^{\tau}ds\, f_{n} (s),
\end{equation}
implying for the memory kernel a factorized expression of the form
\begin{equation}
   \label{eq:25}
   W_{mn} (t)=\pi_{mn}k_n (t),
\end{equation}
where the memory functions $k_m(t)$ relate waiting time
distribution $f_{n} (\tau)$ and survival probability $g_{n}
(\tau)$ through the integral relation
\begin{equation}
   \label{eq:27}
    f_{n} (\tau) =\int_0^\tau ds \, k_n (s)g_{n}(\tau-s)= (k_n \ast
g_{n})(\tau),
\end{equation}
corresponding to Eq.~\eqref{eq:20bis}. Also in this case it is
convenient to express these identities in the Laplace domain, so
that one has
\begin{equation}
   \label{eq:28}
   \hat g_{n} (u)=\frac{1-\hat f_{n} (u)}{u},
\end{equation}
leading to a factorized expression for the memory kernel
\begin{equation}
   \label{eq:29}
   \hat W_{mn} (u) = \pi_{mn} \hat k_{n} (u)
   = \pi_{mn}\frac{\hat f_{n} (u)}{\hat g_{n} (u)},
\end{equation}
which together with Eq.~\eqref{eq:28} yields the following
one-to-one correspondence between $\hat k_{n} (u)$ and $\hat
f_{n}(u)$,
\begin{equation} \label{eq:30}
 \hat k_{n} (u)=\frac{u \hat f_{n} (u)}{1-\hat f_{n} (u)}.
\end{equation}
This relation provides the most direct way to obtain the memory
function $k_n(t)$ given a certain waiting time distribution
$f_n(t)$.

It immediately appears from Eq.~\eqref{eq:25} that the positivity
of the matrix elements of the memory kernel for the considered
class of factorized expressions depends on the positivity of the
memory functions $k_{n} (t)$. We will now consider simple and
natural examples of waiting time distributions leading to negative
memory functions, at variance with what happens in the Markovian
case. To this end the dependence on the index $n$ is not relevant,
since we are only interested in showing that a well-defined waiting time distribution $f (\tau)$
associated to a fixed state of the system can correspond to a
negative function $k (t)$. This point will turn out to be of
particular relevance in the quantum extension of the model, both
in order to identify the class of admissible memory kernels
together with possible pitfalls, and to make contact with relevant
physical models.

Let us consider a general class of waiting time distributions
given by the so-called special Erlang distributions (of order
$a\in \mathbb{N}$)
\begin{equation}
   \label{eq:31}
   f^{(a)} (\tau)=\lambda\frac{(\lambda \tau)^{a-1}}{(a-1)!}e^{-\lambda \tau},
\end{equation}
whose Laplace transform is simply given by
\begin{equation}
   \label{eq:32bis}
   \hat f^{(a)} (u)=\left( \frac{\lambda}{u+\lambda} \right)^{a}.
\end{equation}
Such a distribution describes a random variable given by the sum
of $a$ independent identically distributed exponential random
variables with the same positive parameter $\lambda$. Exploiting
the relation \eqref{eq:30} and inverting the Laplace transform,
which is easily done since we are dealing with rational functions,
one obtains for the first three orders
\begin{align} \label{eq:33}
 f^{(1)} (\tau)&=\lambda e^{-\lambda \tau} & k^{(1)}(t)&= 2\lambda\delta(t)
 \\
 f^{(2)} (\tau)&=\lambda^2 \tau e^{-\lambda \tau}   & k^{(2)} (t)&= \lambda^2
 e^{-2 \lambda t}
 \\
 f^{(3)} (\tau)&=\frac{\lambda^3}{2} \tau^2 e^{-\lambda \tau}   & k^{(3)}
 (t)&= \frac{2\lambda^2}{\sqrt{3}}
 \sin(\sqrt{3\lambda}t/2) e^{-3\lambda t/{2}  },
\end{align}
so that for $a=3$ one indeed has negative contributions in the
memory kernel. On similar grounds one can consider a sum of
exponential random variables characterized by different
parameters, still obtaining a rational function for the Laplace
transform of the waiting time distributions, corresponding to
so-called generalized Erlang distributions. Their expression is
given by
\begin{equation}
   \label{eq:31bis}
   f_{(a)} (\tau)=\sum_{i}^{a}  \left( \prod_{j\not = i}
\frac{\lambda_{j}}{\lambda_{j}-\lambda_{i}} \right)
\lambda_{i}e^{-\lambda_{i} \tau},
\end{equation}
and correspondingly
\begin{equation}
   \label{eq:32}
   \hat f_{(a)} (u)=\prod_{i}^{a}  \frac{\lambda_{i}}{u+\lambda_{i}} .
\end{equation}
In this case for $a=1$ one is obviously back to a simple
exponential distribution, while for  $a=2$ the waiting time
distribution is a difference of two exponential functions
\begin{equation}\label{eq:35}
 f_{(2)}(\tau) = \frac{\lambda_{1}\lambda_{2}}{\lambda_{2}-\lambda_{1}}
(e^{- \lambda_{1} \tau}-e^{-\lambda_{2} \tau})
\end{equation}
leading to the following positive memory function
\begin{equation}
   \label{eq:34}
   k_{(2)} (t)=\lambda_{1}\lambda_{2}e^{- (\lambda_{1}+\lambda_{2}) t}.
\end{equation}
For $a=3$ depending on the value of the three parameters
$\{\lambda_{i}\}_{i=1\ldots 3}$ the memory function can become
oscillatory, thus taking on negative values, according to
\begin{equation}
   \label{eq:36}
   k_{(3)} (t)=\lambda_{1}\lambda_{2}\lambda_{3}
\frac{e^{\lambda_{+} t}-e^{\lambda_{-} t}}{\lambda_{+}-\lambda_{-}},
\end{equation}
with
\begin{equation}\label{eq:37}
 \lambda_{\pm} = -\frac{\lambda_{1}+\lambda_{2}+\lambda_{3}}{2}
 \pm\frac{1}{2}\sqrt{(\lambda_{1}-\lambda_{2}-\lambda_{3})^2-4\lambda_{2}\lambda_{3}}.
\end{equation}

A complementary example is obtained taking rather than a sum of
exponential random variables a single random variable with a
waiting time distribution given by a multi-exponential, that is to
say a convex mixture of exponential distributions
\begin{equation}
   \label{eq:40}
   f (\tau)=\sum_{i}p_{i}\lambda_{i}e^{- \lambda_{i} \tau}, \qquad
   p_{i}\geq 0, \qquad \sum_{i}p_{i}=1.
\end{equation}
Already for the simplest nontrivial case given by a bi-exponential
distribution
\begin{equation}
   \label{eq:38}
   f (\tau)=p \lambda_1 e^{- \lambda_{1} \tau} + (1-p)\lambda_2 e^{- \lambda_{2} \tau},
\end{equation}
with $0< p< 1$, one obtains a memory function taking on negative
values according to
\begin{equation} \label{eq:39}
 k (t) = \langle\lambda\rangle \left[2\delta(t)
 - \frac{\Delta\lambda^2}{\langle\lambda\rangle}
 e^{- (p\lambda_{2}+(1-p)\lambda_1)t}\right],
\end{equation}
where with obvious notation $\langle\lambda\rangle=p \lambda_1 +
(1-p)\lambda_2$ and
$\Delta\lambda^2=\langle\lambda^2\rangle-\langle\lambda\rangle^2$.
It is important to stress that the function given by
Eq.~\eqref{eq:31bis} cannot be interpreted as a multi-exponential
distribution, since the weights in the sum over $i$ are not always
positive. Indeed the situations described by Erlang or
multi-exponential distributions correspond to two complementary
pictures. In both cases the system moves from one state to another
in various unobserved stages or steps, each taking an
exponentially distributed time. In the case of an Erlang
distribution such as Eqs.~\eqref{eq:31} or \eqref{eq:31bis}
however the different steps are taken in series, while for a
multi-exponential distribution as given by Eq.~\eqref{eq:40} the
different stages are entered in parallel, following one of the
available possibilities, each with its own weight. In both cases
one describes a non-Markovian situation in terms of elementary
Markovian building blocks described by exponential distributions,
the non-Markovian features appearing because one does not have
information on the fictitious intermediate steps.  For a suitable
choice of weights and parameters one can approximate any
distribution by combination of stages in series and in parallel,
so that these examples are in fact quite representative.

\section{Quantum semi-Markov processes}\label{sec:quantum-semi-markov}

\subsection{Quantum Markov processes}\label{sec:quantum-markov}
In the Markovian regime the dynamics of the density matrix
$\rho(t)$ of an open quantum system is governed by a master
equation of the relatively simple form of a first-order
differential equation,
\begin{equation} \label{LINDBLAD}
 \frac{d}{dt}\rho(t) = {\mathcal{L}}\rho(t),
\end{equation}
where ${\mathcal{L}}$ is a time-independent infinitesimal
generator with the general structure
\begin{equation} \label{L-STRUCTURE}
 {\mathcal{L}}\rho = -i[H,\rho] +
 \sum_{\alpha} \gamma_{\alpha} \left[ A_{\alpha} \rho A^{\dagger}_{\alpha}
 - \frac{1}{2} \left\{A^{\dagger}_{\alpha}A_{\alpha},\rho\right\} \right].
\end{equation}
The Hamiltonian $H$ describes the coherent part of the time
evolution, while the $A_{\alpha}$ are operators representing the
various decay modes, with $\gamma_{\alpha}\geq 0$ the
corresponding positive decay rates. The solution of
Eq.~(\ref{LINDBLAD}) can be written in terms of a linear map
$V(t)=\exp({\mathcal{L}}t)$ that transforms the initial state
$\rho(0)$ into the state $\rho(t)$ at time $t\geq 0$,
\begin{equation}
 \rho(0) \longrightarrow \rho(t) = V(t)\rho(0).
\end{equation}
The map $V(t)$ is a well-defined quantum dynamical map provided it preserves trace and positivity when applied to a
general initial state $\rho(0)$. This is granted if $V(t)$ is a completely
positive map, in accordance with general
physical principles \cite{Breuer2007}. This property implies that it can be written
in the Kraus form
\begin{equation} \label{KRAUS-REP}
 V(t)\rho(0) =
 \sum_{\alpha}\Omega_{\alpha}(t)\rho(0)\Omega_{\alpha}^{\dagger}(t),
\end{equation}
where in order to grant preservation of the trace the operators $\Omega_{\alpha}(t)$ have the property that
the sum
$\sum_{\alpha}\Omega_{\alpha}^{\dagger}(t)\Omega_{\alpha}(t)$ is
equal to the unit operator. Hence, $V(t)$ represents a completely
positive dynamical semigroup known as quantum Markov process. The master equation (\ref{LINDBLAD}) leads to
such a semigroup if and only if the generator is of the form of
Eq.~(\ref{L-STRUCTURE}). This is the content of the celebrated
Gorini-Kossakowski-Sudarshan-Lindblad theorem
\cite{Gorini1976a,Lindblad1976a}, of paramount importance  in both
fundamental and phenomenological approaches to the description of
irreversible dynamics in quantum mechanics \cite{Breuer2007}.

For the case in which one has a closed system of equations for the
populations $P_n (t)=\langle n|\rho (t)|n\rangle$ in a fixed
orthonormal basis one recovers from Eq.~\eqref{LINDBLAD} and
Eq.\eqref{L-STRUCTURE} the Pauli master equation (\ref{eq:23}). This
justifies the notion of a quantum Markov process and provides a direct
connection to a classical Markov process.

\subsection{Master equations with memory kernel}\label{sec:q-memory-kernel}
A natural non-Markovian generalization of Eq.~(\ref{LINDBLAD})
is given by the integrodifferential equation
\begin{equation}\label{NOnon-MarkovianARKOV}
 \frac{d}{dt}\rho(t) = \int_0^t d\tau \, {\mathcal{K}}(\tau) \rho(t-\tau).
\end{equation}
Here quantum memory effects are taken into account
through the introduction of the memory kernel
${\mathcal{K}}(\tau)$, which means that the rate of change of the
state $\rho(t)$ at time $t$ depends on the states $\rho(t-\tau)$
at previous times $t-\tau$. Equations of the form
(\ref{NOnon-MarkovianARKOV}) typically arise in the
standard Nakajima-Zwanzig projection operator technique
\cite{Nakajima1958a,Zwanzig1960a}. As an important limiting case, the Markovian master
equation (\ref{LINDBLAD}) is recovered for a memory kernel proportional to a $\delta$-function,
\begin{equation}\label{M-LIMIT}
 {\mathcal{K}}(\tau)=2\delta(\tau){\mathcal{L}}.
\end{equation}

To be physically acceptable the superoperator $\mathcal{K}(\tau)$
appearing in Eq.~\eqref{NOnon-MarkovianARKOV} must lead to a
completely positive quantum dynamical map $V(t)$. The general
structural characterization of the memory kernels with this
property is an unsolved problem of central importance in the field
of non-Markovian quantum dynamics
\cite{Budini2004a,Daffer2004a,Shabani2005a,Wilkie2009a}. In fact,
even the most simple and natural choices for $\mathcal{K}(\tau)$
can lead to unphysical results \cite{Barnett2001a,Budini2004a}.
Here we will construct a class of memory kernels which arises
naturally as a quantum mechanical generalization of the classical
semi-Markov processes, and allows the formulation of criteria that
guarantee complete positivity.

We consider memory kernels with the general structure
\begin{eqnarray}
   \label{eq:1q}
   \mathcal{K} (\tau)\rho&=& -i\left[H(\tau),\rho\right]
 - \frac 12 \sum_{\alpha} \gamma_{\alpha}(\tau)
 \left\{A^{\dagger}_{\alpha}(\tau)A_{\alpha}(\tau),\rho\right\}
 \nonumber \\
 &~& +\sum_{\alpha} \gamma_{\alpha}(\tau) A_{\alpha}(\tau)\rho
 A^{\dagger}_{\alpha}(\tau),
\end{eqnarray}
that is to say of the form given by Eq.~\eqref{L-STRUCTURE} apart
from the time dependence of the operators $A_{\alpha}(\tau)$ and
of the real functions $\gamma_{\alpha}(\tau)$. As previously done
in the Markovian case let us consider the situation in which the
populations obey a closed system of equations of motion, which
then takes the form (\ref{eq:20}) of the generalized master
equation for a classical semi-Markov process, where the memory
kernel is given by
\begin{equation}
 W_{nm}(\tau)=\sum_\alpha \gamma_{\alpha}(\tau)
 |\langle n |A_{\alpha}(\tau)|m\rangle|^2.
\end{equation}
Thus, whenever the populations obey closed equations,
Eq.~\eqref{LINDBLAD} yields the classical Markovian master
equation \eqref{eq:23}, while Eq.~\eqref{NOnon-MarkovianARKOV}
with the kernel \eqref{eq:1q} leads under the same conditions to
the generalized master equation (\ref{eq:20}) for a classical
semi-Markov process. This justifies
the name quantum semi-Markov process.

\subsection{Conditions for complete positivity}\label{sec:CP-conditions}
Our next goal is the formulation of sufficient conditions that
guarantee the complete positivity of the dynamical map $V(t)$
corresponding to the non-Markovian master equation defined by
Eqs.~(\ref{NOnon-MarkovianARKOV}) and \eqref{eq:1q}, no longer
assuming that closed equations for the populations exist.

\subsubsection{Quantum dynamical map}
The dynamical map $V(t)$ corresponding to the master equation
(\ref{NOnon-MarkovianARKOV}) is defined as the solution of the
integrodifferential equation
\begin{equation} \label{VT}
 \frac{d}{dt}V(t) = \int_0^t d\tau \, {\mathcal{K}}(\tau) V(t-\tau)
\end{equation}
with the initial condition $V(0)=I$, where $I$ denotes the
identity map. Following Ref.~\cite{Kossakowski2008a} let us decompose
the memory kernel as
\begin{equation} \label{DECOMP}
 {\mathcal{K}}(\tau) = B(\tau) + C(\tau),
\end{equation}
where the superoperators $B(\tau)$ and $C(\tau)$ are defined by
\begin{eqnarray}
 B(\tau)\rho &=& \sum_{\alpha} \gamma_{\alpha}(\tau)
 A_{\alpha}(\tau)\rho A^{\dagger}_{\alpha}(\tau),
 \label{B-DEF} \\
 C(\tau)\rho &=& -i\left[H(\tau),\rho\right] \nonumber \\
 && - \frac 12 \sum_{\alpha}
 \gamma_{\alpha}(\tau)\left\{A^{\dagger}_{\alpha}(\tau)A_{\alpha}(\tau),\rho\right\}.
 \label{C-DEF}
\end{eqnarray}
We further introduce the map $V_0(t)$ as the solution of the
equation
\begin{equation} \label{R-EQ}
 \frac{d}{dt}V_0(t) = \int_0^t d\tau\, C(\tau) V_0(t-\tau),
\end{equation}
with the initial condition $V_0(0)=I$. The Laplace transformation
of Eqs.~(\ref{VT}) and (\ref{R-EQ}) yields
\begin{equation} \label{LAPLACE-TRAFO}
 \hat{V}(u) = \frac{1}{u-\hat{\mathcal{K}}(u)}, \qquad
 \hat{V}_0(u) = \frac{1}{u-\hat{C}(u)},
\end{equation}
from which we get the Dyson-type identity
\begin{equation}
 \hat{V}(u) = \hat{V}_0(u) + \hat{V}_0(u) \hat{B}(u) \hat{V}(u).
\end{equation}
Transforming back to the time domain we obtain the equation
\begin{equation} \label{DYSON}
 V(t) = V_0(t) + (V_0 \ast B \ast V)(t).
\end{equation}
Regarding formally the superoperator $B(\tau)$ as a perturbation
and iterating Eq.~(\ref{DYSON}) one finds that the full dynamical
map $V(t)$ can be represented as a series,
\begin{eqnarray} \label{V-REP}
 V(t) &=& V_0(t) + (V_0\ast B\ast V_0)(t) \nonumber \\
 &~& + (V_0\ast B\ast V_0\ast B \ast V_0)(t) + \ldots,
\end{eqnarray}
which turns out to be a useful relation in the formulation of
appropriate conditions for complete positivity.

\subsubsection{Sufficient conditions for complete
               positivity}\label{sec:SUFF-COND}
Let us first assume that the quantities $\gamma_{\alpha}(\tau)$
are positive functions, which means that the superoperator
$B(\tau)$ defined by Eq.~(\ref{B-DEF}) is completely positive.
Since the property of the complete positivity is preserved under
addition and convolution, the representation (\ref{V-REP}) then
tells us that the full dynamical map $V(t)$ is completely positive
if the map $V_0(t)$ is completely positive. To bring this
condition into an explicit form let us assume further that the
Hermitian operators $H(\tau)$ and
$\sum_{\alpha}\gamma_{\alpha}(\tau)A^{\dagger}_{\alpha}(\tau)A_{\alpha}(\tau)$
are diagonal in a time-independent orthonormal basis
$\{|n\rangle\}$ for the underlying Hilbert space, i.~e. we have
\begin{eqnarray}
 H(\tau) &=& \sum_n \varepsilon_n(\tau) |n\rangle\langle n|, \label{H-TAU} \\
 \sum_{\alpha} \gamma_{\alpha}(\tau) A^{\dagger}_{\alpha}(\tau)A_{\alpha}(\tau)
 &=& \sum_n k_n(\tau) |n\rangle\langle n|, \label{AA-TAU}
\end{eqnarray}
with in general time-dependent eigenvalues $\varepsilon_n(\tau)$
and $k_n(\tau)$. Note that the positivity of the
$\gamma_{\alpha}(\tau)$ implies that the eigenvalues $k_n(\tau)$
must be positive as well.

Equation (\ref{R-EQ}) can now be solved to obtain
\begin{equation} \label{R-REP}
 V_0(t)\rho (0)  = \sum_{nm} g_{nm}(t)
 |n\rangle\langle n|\rho (0) |m\rangle\langle m|,
\end{equation}
where the functions $g_{nm}(t)$ are the solutions of
\begin{equation} \label{Gnon-Markovian}
 \dot{g}_{nm}(t) = -\int_0^t d\tau
 \left[ z_n(\tau)+z_m^{\ast}(\tau) \right] g_{nm}(t-\tau),
\end{equation}
corresponding to the initial conditions $g_{nm}(0)=1$, and
\begin{equation}
 z_n(\tau)=\frac{1}{2}k_n(\tau)+i\varepsilon_n(\tau).
\end{equation}
Equation (\ref{R-REP}) can be proven as follows. First, one shows
that $C(\tau)\left(|n\rangle\langle
m|\right)=-[z_n(\tau)+z_m^{\ast}(\tau)]|n\rangle\langle m|$. Using
this relation and differentiating Eq.~(\ref{R-EQ}) one easily
demonstrates that the expression (\ref{R-REP}) indeed represents
the desired solution.

It is important to notice that the functions $g_{nn} (t)$ do
actually coincide with the survival probabilities $g_{n}(t)$
introduced in Eq.~\eqref{eq:24}. In fact, for $n=m$ we get from
Eq.~(\ref{Gnon-Markovian})
\begin{equation}\label{g_nn}
 \dot{g}_{nn}(t) = -\int_0^t d\tau k_n(\tau) g_{nn}(t-\tau).
\end{equation}
The survival probabilities satisfy the same equation as can be
seen by taking the time derivative of Eq.~(\ref{eq:24}) and using
Eq.~(\ref{eq:27}).

One easily verifies that the representation (\ref{R-REP}) can be
brought into the Kraus form (\ref{KRAUS-REP}) if and only if the
matrix $G(t)$ with the elements $g_{nm}(t)$ is positive. Thus, we
arrive at a sufficient condition for complete positivity: The
quantum dynamical map $V(t)$ corresponding to the non-Markovian
master equation (\ref{NOnon-MarkovianARKOV}) with the memory
kernel \eqref{eq:1q} is completely positive if the condition
\begin{equation} \label{COND-1}
 G(t) = (g_{nm}(t)) \geq 0
\end{equation}
is fulfilled.

A necessary condition for \eqref{COND-1} to hold is the positivity
of the diagonal elements $g_{nn}(t)$ of $G (t)$ which coincide
with the survival probabilities, $g_n(t)=g_{nn}(t)$. This
necessary condition in turn implies the positivity of the
functions $f_{n}(t)$ as can be seen immediately from
Eq.~\eqref{eq:27} because $k_n(t)\geq 0$. Condition (\ref{COND-1})
therefore implies that the functions $f_n(t)$ allow an
interpretation as true waiting time distributions. The positivity
of the matrix $G(t)$ therefore represents a natural quantum
generalization of the classical conditions for a semi-Markov
process.

\subsubsection{Markovian limit and Lindblad theorem}
In the Markovian limit expressed by Eq.~(\ref{M-LIMIT}) we must
have $k_n(\tau)=2k_n^0\delta(\tau)$ and
$\varepsilon_n(\tau)=2\varepsilon_n^0\delta(\tau)$, such that
$z_n(\tau)=2z_n^0\delta(\tau)$. Equation (\ref{Gnon-Markovian})
then reduces to the time-local equation
\begin{equation}
 \dot{g}_{nm}(t) = -\left(z_n^0+z_m^{0\ast}\right) g_{nm}(t),
\end{equation}
which is easily solved to yield
\begin{equation}
 g_{nm}(t) = h_n(t) h_m^{\ast}(t), \qquad h_n(t) = e^{-z_n^0t}.
\end{equation}
Since a matrix with elements of this form is always positive we
conclude that condition (\ref{COND-1}) is automatically satisfied.
It follows from our results that the corresponding quantum
dynamical map represents a completely positive semigroup. Hence,
we see that our formulation correctly describes the transition to
the Markovian limit and that it contains as a special case the
``if''-part of the Lindblad theorem.

\subsubsection{Negative memory functions}\label{sec:NEGATIVE-MEMORY}
Up to now we have considered the case that the superoperator
$B(\tau)$ is completely positive which led to the sufficient
condition (\ref{COND-1}) for the complete positivity of the
dynamical map. The complete positivity of $B(\tau)$ implies that
the memory functions must be positive, $k_n(t)\geq 0$. However, in
many physical relevant applications these functions can take on
negative values.

To include cases with negative kernel functions we consider the
following general class of memory kernels. We use again the
decomposition of the form given by Eq.~(\ref{DECOMP}), where
$C(\tau)$ is given by the expressions (\ref{C-DEF}), (\ref{H-TAU})
and (\ref{AA-TAU}). However, we now drop the condition that the
map $B(\tau)$ is completely positive, supposing instead that it
takes the following general form,
\begin{equation} \label{DEF-B-NEW}
 B(\tau) = \sum_{n} k_n(\tau) B_n,
\end{equation}
where the $k_n(\tau)$ are real functions, not necessarily
positive, and the $B_n$ are completely positive and
time-independent maps. The full memory kernel can thus be written
as
\begin{equation}
 {\mathcal{K}}(\tau)\rho = -i[H(\tau),\rho] + \sum_n k_n(\tau)
 \left( B_n\rho - \frac{1}{2} \left\{|n\rangle\langle n|,\rho\right\}\right).
\end{equation}

Without the assumption of the complete positivity of $B(\tau)$ the
complete positivity of $V_0(t)$ is generally not sufficient for
the complete positivity of the dynamical map $V(t)$. However, we
can conclude from the representation (\ref{V-REP}) that $V(t)$ is
completely positive if both $V_0(t)$ and $(V_0\ast B)(t)$ are
completely positive. We have
\begin{equation}\label{V0-B}
 (V_0\ast B)(t)\rho = \sum_{lnm} f_{nm}^l(t)
 |n\rangle\langle n|B_l\rho|m\rangle\langle m|,
\end{equation}
where
\begin{equation}
 f^l_{nm}(t) = \int_0^t d\tau k_{\,l}(\tau) g_{nm}(t-\tau).
\end{equation}
The map (\ref{V0-B}) is completely positive if for all $l$ the
matrix with elements $f^l_{nm}(t)$ is positive,
\begin{equation} \label{COND-2}
 F^l(t) = (f^l_{nm}(t)) \geq 0.
\end{equation}
Summarizing, we have shown that the quantum dynamical map $V(t)$
is completely positive if the conditions (\ref{COND-1}) and
(\ref{COND-2}) are fulfilled.

We can again provide the connection of the obtained results to the
interpretation in terms a classical semi-Markov process. In fact,
a necessary condition for (\ref{COND-2}) is the positivity of the
functions $f^n_{nn}(t)$ which coincide with the functions
$f_n(t)$. Conditions (\ref{COND-1}) and (\ref{COND-2}) thus imply
that $g_n(t)$ and $f_n(t)$ can be interpreted as survival
probability and as waiting time distribution for a classical
semi-Markov process, respectively. These conditions therefore
represent a generalization of the classical conditions to the
quantum case for memory functions $k_n(t)$ that are allowed to
take on negative values.

\subsection{Examples}\label{sec:examples}
To illustrate the theory developed here and the various conditions
for the complete positivity of the dynamical map we will now
introduce a few examples. In particular we will consider a
structure for the memory kernel which encompasses and generalizes
a model recently studied in the literature for the description of
memory effects \cite{Budini2004a,Maniscalco2007a}. For this class
of memory kernels we are in particular able to give necessary and
sufficient conditions for the complete positivity of the dynamical
evolution.

\subsubsection{Lattice systems}
\label{sec:n-level-system} Let us consider the following memory
kernel,
\begin{eqnarray} \label{QSM}
 {\mathcal{K}}(\tau)\rho  &=&
 -i\left[H(\tau),\rho \right] -\frac 12 \sum_n
 k_n(\tau) \left\{|n\rangle\langle n|,\rho \right\}, \nonumber \\
 &~& + \sum_{mn}
 \pi_{mn} k_n(\tau) |m\rangle\langle n|\rho |n\rangle\langle m|.
\end{eqnarray}
The special feature of this kernel is given by the fact that it
leads to closed equations of motion for the populations
$P_n(t)=\rho_{nn}(t)=\langle n|\rho(t)|n\rangle$ and for the
coherences $\rho_{nm}(t)=\langle n|\rho(t)|m\rangle$, $n\neq m$.
In fact, we find from the master equation with the above kernel
that the coherences satisfy the same equation as the quantities
$g_{nm}(t)$, namely Eq.~(\ref{Gnon-Markovian}). Taking into
account the initial conditions we thus have
\begin{equation}\label{REPR-1}
 \rho_{nm}(t) = g_{nm}(t) \rho_{nm}(0), \qquad n \neq m.
\end{equation}
On the other hand, the populations are found to obey the
generalized master equation (\ref{eq:20}) with
$W_{nm}(\tau)=\pi_{nm}k_m(\tau)$. We assume that this master
equation describes a classical semi-Markov process and denote by
$T_{nm}(t)$ the corresponding conditional transition probability.
The diagonals of the density matrix can therefore be written as
\begin{equation}\label{REPR-2}
 \rho_{nn}(t) = \sum_m T_{nm}(t) \rho_{mm}(0).
\end{equation}
Thus, the memory kernel (\ref{QSM}) may be viewed as describing a
quantum particle moving on a lattice with sites labelled by $n$.
The dynamics of the populations is modelled through a semi-Markov
process with transition probabilities $\pi_{nm}$ and arbitrary
waiting time distributions $f_n(t)$, while the $\rho_{nm}(t)$
describe the quantum coherences between different sites $n$ and
$m$.

With the help of Eqs.~(\ref{REPR-1}) and (\ref{REPR-2}) we can
immediately construct the quantum dynamical map,
\begin{eqnarray}\label{REPR-3}
 V(t)\rho(0) &=& \sum_{n\neq m} g_{nm}(t)
 |n\rangle\langle n|\rho(0)|m\rangle\langle m| \nonumber \\
 && + \sum_{nm} T_{nm}(t) |n\rangle\langle m|\rho(0)|m\rangle\langle n|.
\end{eqnarray}
Let us introduce a matrix $\tilde{G}(t)=(\tilde{g}_{nm}(t))$ whose
elements are defined by
\begin{equation}\label{TILDE-G}
 \tilde{g}_{nm}(t) = (T_{nn}(t)-g_{nn}(t))\delta_{nm} + g_{nm}(t).
\end{equation}
The off-diagonal elements of $\tilde{G}(t)$ thus coincide with
those of the matrix $G(t)$ introduced earlier, while the diagonals
of $\tilde{G}(t)$ are given by the conditional transition
probabilities $T_{nn}(t)$. Then we can rewrite Eq.~(\ref{REPR-3})
as
\begin{eqnarray}\label{REPR-4}
 V(t)\rho(0) &=& \sum_{nm} \tilde{g}_{nm}(t)
 |n\rangle\langle n|\rho(0)|m\rangle\langle m| \nonumber \\
 && + \sum_{n\neq m} T_{nm}(t) |n\rangle\langle m|\rho(0)|m\rangle\langle n|.
\end{eqnarray}
This is an exact formal representation for the full quantum
dynamical map from which we infer that $V(t)$ is completely
positive if and only if $\tilde{G}(t)\geq 0$ and $T_{nm}(t)\geq 0$
for all $n\neq m$. Being the conditional transition probabilities
of a semi-Markov process, the $T_{nm}(t)$ always satisfy of course
the second condition. Hence we obtain the result that the quantum
dynamical map $V(t)$ is completely positive if and only if the
condition
\begin{equation} \label{COND-3}
 \tilde{G}(t) = (\tilde{g}_{nm}(t)) \geq 0
\end{equation}
holds. This condition provides a full characterization of the
complete positivity of the class of quantum semi-Markov processes
given by Eq.~\eqref{QSM}.

Assuming the memory functions to be positive, the kernel
(\ref{QSM}) is easily seen to be of the form introduced in
Sec.~\ref{sec:SUFF-COND} with the superoperator $B(\tau)$ given by
the completely positive map
\begin{equation}
 B(\tau)\rho = \sum_{mn} \pi_{mn} k_n(\tau)
 |m\rangle\langle n|\rho|n\rangle\langle m|.
\end{equation}
Hence we can apply condition (\ref{COND-1}) as sufficient
condition for the complete positivity of $V(t)$. We note that the
probabilities $T_{nn}(t)$ are in general larger than the
corresponding survival probabilities $g_n(t)=g_{nn}(t)$, since the
process can be in the initial state $n$ at time $t$ both because
it has not left it and because it has come back to it. According
to Eq.~(\ref{TILDE-G}) the necessary and sufficient condition
(\ref{COND-3}) is therefore in general weaker than the merely
sufficient condition (\ref{COND-1}). However, if the process
involves only jumps in one specific direction, i.~e., if the
return probability vanishes for all states (this happens, e.~g., for
a purely decaying system), we have $T_{nn}(t)=g_n(t)$ and, hence,
condition (\ref{COND-1}) becomes a necessary and sufficient
condition for the complete positivity.

An instructive special case is that of a translational invariant
system for which $k_n(\tau)$, $g_{nm}(t)$ and $T_{nn}(t)$ are
state-independent, i.~e. $k_n(\tau)=k(\tau)$, $g_{nm}(t)=g(t)$ and
$T_{nn}(t)=T(t)$. The conditions (\ref{COND-1}) and (\ref{COND-3})
are then automatically fulfilled, showing that any such
translational invariant process leads to a completely positive
dynamical map. The same conclusion holds true if we allow the
memory function $k(\tau)$ to become negative. In fact, the kernel
(\ref{QSM}) is also of the form used in
Sec.~\ref{sec:NEGATIVE-MEMORY} with $B_n(\tau)$ given by
\begin{equation}
 B_n\rho = \sum_m \pi_{mn} |m\rangle\langle n|\rho|n\rangle\langle m|.
\end{equation}
Condition (\ref{COND-2}) is then also satisfied by assumption
because it reduces to the condition that the waiting time
distribution $f(\tau)$ corresponding to the memory function
$k(\tau)$ must be positive.

\subsubsection{Exponential memory functions}\label{sec:expon-memory-kern}
Memory kernels of the form \eqref{QSM} describe e.g. the dynamics
of a two-level system interacting with a bosonic quantum
reservoir, such as for example a two-level atom coupled to a
damped field mode. The index $n$ now only takes on the two values
$+$ and $-$ and the memory kernel reads
\begin{eqnarray}
   \label{eq:41}
    \mathcal{K}(\tau) \rho &=&
 k_{+}(\tau) \left[ \sigma_-\rho \sigma_+
 - \frac{1}{2} \left\{ \sigma_+\sigma_-,\rho \right\} \right] \nonumber \\
 &~& + k_{-}(\tau) \left[ \sigma_+\rho \sigma_-
 - \frac{1}{2} \left\{ \sigma_-\sigma_+,\rho \right\} \right],
\end{eqnarray}
where the jump probabilities are given by $\pi_{+-}=\pi_{-+}=1$
and $\pi_{--}=\pi_{++}=0$. A typical expression arising for the
time dependence of the memory functions is given by
\cite{Breuer2007,Budini2004a,Maniscalco2007a}
\begin{equation}
   \label{eq:42}
   k_{\pm}(\tau) =\kappa_{\pm}e^{-\gamma \tau},
\end{equation}
with decay constants $\kappa_{\pm}\geq 0$. This expression also
allows an explicit evaluation of the relevant quantities such as
the functions $g_{nm}(t)$ and the conditional transition
probabilities $T_{nm}(t)$. Inverting Eq.~\eqref{eq:30} to obtain
\begin{equation}
   \label{eq:43}
   \hat f_{\pm} (u)=\frac{\hat k_{\pm} (u)}{u+\hat k_{\pm} (u)},
\end{equation}
and further calculating the inverse Laplace transform one finds
the functions
\begin{equation}
   \label{eq:44}
   f_{\pm} (\tau)=
   2 \frac{\kappa_{\pm}}{d_{\pm}} e^{-\gamma  \tau/2}\sinh (d_{\pm}\tau/2)
\end{equation}
with
\begin{equation}
   \label{eq:52}
   d_{\pm}=\sqrt{\gamma^2-4\kappa_{\pm}}.
\end{equation}
The functions $f_{\pm} (\tau)$ are positive and normalized to one (or
identically zero) if and only if
\begin{equation}
   \label{eq:51}
   \frac{\gamma^2}{4}\geq \max\{\kappa_{+},\kappa_{-}\},
\end{equation}so that only in this case they can be
interpreted as waiting time distributions and the generalized master
equation corresponding to Eq.~\eqref{eq:41} describes a classical
semi-Markov process. In fact under the condition $\gamma^2\geq
4\kappa_+$ the function $f_{+} (\tau)$ can be expressed as a
difference of two exponentials as in \eqref{eq:35}
with the identifications
\begin{equation}
   \label{eq:45}
   \lambda_{1,2}=\frac{\gamma}{2}\pm\frac 12 \sqrt{\gamma^2-
4\kappa_+},
\end{equation}
thus corresponding to a generalized Erlang distribution, and
similarly for the function $f_{-} (\tau)$.

The conditional transition probabilities for the associated
semi-Markov process can be calculated observing that due to
Eq.~\eqref{REPR-2} the quantity $T_{++} (t)$ is given by the
solution $\rho_{++} (t)$ of the master equation obtained with the
initial condition $\rho_{++} (0)=1$ and similarly for $T_{--}
(t)$.  Starting from Eq.~\eqref{eq:41} one has the equation
\begin{equation}
   \label{eq:46}
   \dot{\rho}_{++}(t)
   =\int_{0}^{t}\left[k_{-}(t-\tau)\rho_{--} (\tau) -k_{+}(t-\tau) \rho_{++}
   (\tau)\right].
\end{equation}
Differentiating this equation with respect to time and exploiting
the exponential form of the memory functions \eqref{eq:42}
together with the trace preservation one is led to a telegraph
equation of the form
\begin{equation}
   \label{eq:47}
 \ddot{\rho}_{++}(t)+\gamma\dot{\rho}_{++}(t)
 +(\kappa_{+}+\kappa_{-}){\rho}_{++}(t)-\kappa_{-}
 = 0.
\end{equation}
Its solution with the initial conditions $\rho_{++} (0)=1$ and
$\dot{\rho}_{++} (0)=0$ gives the conditional transition
probability $T_{++} (t)$, and an analogous calculation can be
performed for $T_{--} (t)$. The results read
\begin{align}
   \label{eq:49}
   T_{++} (t) =&
   \frac{\kappa_{-}}{\kappa_{+}+\kappa_{-}}
\nonumber\\
& +\frac{\kappa_{+}}{\kappa_{+}+\kappa_{-}} e^{-\gamma t/2}
\!
\left[
\cosh(dt/2) + \frac{\gamma}{d}
 \sinh(dt/2)\right],
\nonumber\\
   T_{--} (t) =&
   \frac{\kappa_{+}}{\kappa_{+}+\kappa_{-}}
\nonumber\\
& +\frac{\kappa_{-}}{\kappa_{+}+\kappa_{-}} e^{-\gamma t/2}
\!
\left[
\cosh(dt/2) + \frac{\gamma}{d}
 \sinh(dt/2)\right]
\end{align}
with
\begin{equation}
   \label{eq:50}
   d=\sqrt{\gamma^2-4 (\kappa_{+}+\kappa_{-})}.
\end{equation}
Note that despite the fact that $d$ is not necessarily real, since
this is generally not implied by Eq.~\eqref{eq:51}, the transition
probabilities themselves are always positive.

For the memory functions \eqref{eq:42} one can also exactly
calculate the entries of the matrix $G (t)$ considered in
\eqref{COND-1}. In fact both $g_{++} (t)$ and $g_{--} (t)$ due to
\eqref{g_nn} and the exponential form of the kernel have a second
derivative which obeys a telegraph equation, and similarly for
$g_{+-} (t)$. The corresponding solutions read
\begin{align}
   \label{eq:53}
      g_{++} (t) &=
e^{-\gamma t/2}\left[ \cosh(d_{+}t/2) + \frac{\gamma}{d_{+}}
 \sinh(d_{+}t/2)\right]
\nonumber\\
      g_{--} (t) &=
e^{-\gamma t/2}\left[ \cosh(d_{-}t/2) + \frac{\gamma}{d_{-}}
 \sinh(d_{-}t/2)\right]
\intertext{and}
      g_{+-} (t) &=
e^{-\gamma t/2}\left[ \cosh(\bar{d}t/2) + \frac{\gamma}{\bar{d}}
 \sinh(\bar{d}t/2)\right],
\end{align}
with $d_{\pm}$ as in Eq.~\eqref{eq:52} while
\begin{equation}
   \label{eq:54}
    \bar{d}=\sqrt{\gamma^2-2 (\kappa_{+}+\kappa_{-})}.
\end{equation}
In particular $g_{+-} (t)=g_{-+} (t)$ since it is a real quantity.
Note that indeed minus the derivative of $g_{++}(t)$ is equal to
$f_{+}(t)$, according to its interpretation as survival
probability, and similarly for $g_{--}(t)$ and $f_{-}(t)$. This
entails in particular that both $g_{++}(t)$ and $g_{--}(t)$ are
positive non-increasing functions taking the value one for $t=0$.

One can check that as discussed in Sec.~\ref{sec:n-level-system}
indeed the inequality $T_{nn}(t)\geq g_{nn}(t)$ generally holds
with $n=\pm$. In fact the quantity $T_{nn}(t)- g_{nn}(t)$ is equal
to zero for $t=0$ and has a non-negative derivative with respect
to time for all values of the parameters complying with
Eq.~\eqref{eq:51}. This is consistent with the interpretation of
the $T_{nn}(t)$ as conditional transition probabilities which must be
larger or equal to the corresponding survival probabilities. In
particular one immediately sees from the explicit expressions
\eqref{eq:49} and \eqref{eq:53} that for $\kappa_{-}=0$, so that
one only has transitions in one direction, $T_{++}(t)=g_{++}(t)$
and $T_{--}(t)=g_{--}(t)=1$.

The necessary and sufficient conditions for complete positivity of
the dynamics described by Eq.~\eqref{eq:41} according to the
general result Eq.~\eqref{COND-3} are thus given by
\begin{equation}
   \label{eq:48}
   T_{++}(t) \; T_{--}(t)\geq g_{-+}^2 (t),
\end{equation}
together with the constraint \eqref{eq:51}, which ensures that the
populations obey a generalized master equation describing a
classical semi-Markov process. One therefore has to find out the
possible range of parameters $\kappa_{+},\kappa_{-}$ and $\gamma$
for which the inequality Eq.~\eqref{eq:48} is satisfied for all
times, or at least up to a certain time. The task of finding
conditions for complete positivity of the dynamics given by
Eq.~\eqref{eq:41} is indeed far from trivial and has been
accomplished only partially in the literature
\cite{Budini2004a,Maniscalco2007a}.

\begin{figure}[htb]
\begin{center}
\includegraphics[width=0.8\linewidth]{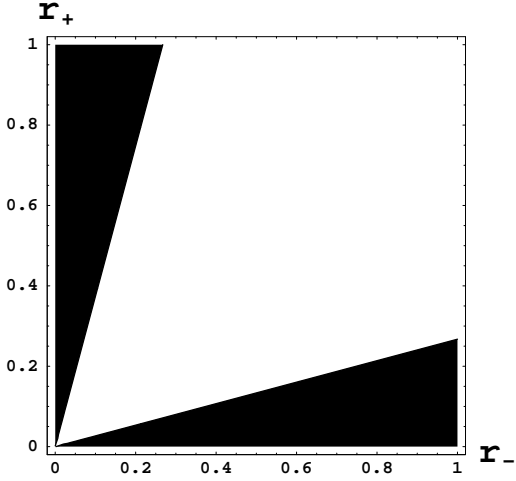}
\caption{The sign of $\Delta(\tau)$ plotted for $\tau=0.01$.
Within the white region $\Delta$ is positive, while it is negative
in the black regions. The range of $r_{\pm}$ is restricted to
$[0,1]$ according to the constraint \eqref{eq:51}.
\label{negative-region}}
\end{center}
\end{figure}

Taking the explicit expressions \eqref{eq:49} and \eqref{eq:53}
into account and noting that for $t=0$ Eq.~\eqref{eq:48} is
actually an equality, one can look at the short-time behavior of
the quantity
\begin{equation}
   \label{eq:55}
   \Delta(t) =T_{++}(t) \; T_{--}(t)- g_{-+}^2 (t).
\end{equation}
In view of the constraint Eq.~\eqref{eq:51} the function $\Delta$
is most conveniently expressed in terms of the rescaled quantities
\begin{equation}
   \label{eq:60}
   r_{\pm}=\frac{4}{\gamma^2}\kappa_{\pm}
\end{equation}
bound to the interval $[0,1]$, and $\tau=\gamma t$. In terms of
these new variables $\Delta$ can be written as
\begin{eqnarray}
 \Delta(\tau) &=&
 \left(\frac{r_-}{r_-+r_+}+\frac{r_+}{r_-+r_+}e^{-\tau/2}h_1(\tau)\right)
 \nonumber \\
 &~& \times \left(\frac{r_+}{r_-+r_+}+\frac{r_-}{r_-+r_+}e^{-\tau/2}h_1(\tau)\right)
 \nonumber \\
 &~& - e^{-\tau}h^2_2(\tau),
\end{eqnarray}
where
\begin{eqnarray}
 h_1(\tau) &=& \cosh\left(\frac{\tau}{2}\sqrt{1-(r_-+r_+)}\right)
 \nonumber \\
 &~& + \frac{\sinh\left(\frac{\tau}{2}\sqrt{1-(r_-+r_+)}\right)}{\sqrt{1-(r_-+r_+)}}
\end{eqnarray}
and
\begin{eqnarray}
 h_2(\tau) &=& \cosh\left(\frac{\tau}{2}\sqrt{1-(r_-+r_+)/2}\right)
 \nonumber \\
 &~& +
 \frac{\sinh\left(\frac{\tau}{2}\sqrt{1-(r_-+r_+)/2}\right)}{\sqrt{1-(r_-+r_+)/2}}.
\end{eqnarray}
The Taylor expansion of $\Delta$ for small $\tau$ reads
\begin{equation}
   \label{eq:56}
   \Delta (\tau)=-\frac{1}{96} (r_{+}^2+r_{-}^2-4
   r_{+}r_{-})\tau^3 +\mathcal{O} (\tau^4),
\end{equation}
so that whenever
\begin{equation} \label{eq:ineq}
 r_{+}^2+r_{-}^2-4 r_{+}r_{-} > 0
\end{equation}
one has violation of complete positivity for very short times.
This is obviously the case for either $r_{-}=0$, $r_+>0$ or
$r_{+}=0$, $r_->0$. The function $\Delta$ is plotted in
Fig.~\ref{negative-region} in the region allowed by
Eq.~\eqref{eq:51}, still clearly taking on negative values. Note
that in standard physical situations one has $r_{+}\geq r_{-}$,
which corresponds to positivity of the reservoir temperature.
Indeed the condition $r_{-}=0$ would correspond to a zero
temperature reservoir. This remark confirms a result obtained in a
completely different way in \cite{Maniscalco2007a}.

\begin{figure}[htb]
\begin{center}
\includegraphics[width=0.8\linewidth]{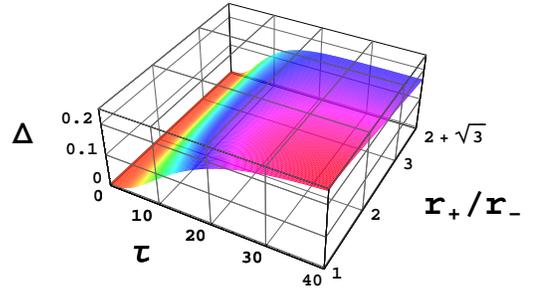}
\caption{(Color online) The quantity $\Delta$ plotted as a
function of $\tau$ and of the ratio between the decay constants in
the range $[1,2+\sqrt{3}]$. $r_{-}$ is fixed to be 0.2.
\label{positive-region}}
\end{center}
\end{figure}

For $r_->0$ the inequality \eqref{eq:ineq} can be rewritten as
\begin{equation}
   \label{eq:57}
   \left (\frac{r_{+}}{r_{-}}\right)^2 -4 \left (\frac{r_{+}}{r_{-}}\right)+1
   > 0,
\end{equation}
which is satisfied whenever
\begin{equation}
   \label{eq:58}
   r_{+}< (2-\sqrt{3})r_{-}\qquad \mathrm{or}\qquad r_{+} > (2+\sqrt{3})r_{-}.
\end{equation}
In these parameter regions, even when the classical condition
Eq.~\eqref{eq:51} holds, the necessary and sufficient condition
Eq.~\eqref{eq:48} for complete positivity is violated for very
short times. Assuming as discussed above $r_{+}\geq r_{-}$ we have
thus obtained that the decay constants must satisfy the constraint
\begin{equation}
   \label{eq:59}
   r_{-}\leq r_{+}\leq (2+\sqrt{3})r_{-},
\end{equation}
so as to avoid loss of complete positivity for very short times. A
numerical analysis indicates that in this parameter region the
inequality Eq.~\eqref{eq:48} is satisfied, corresponding to
preservation of complete positivity. Indeed the triangular white
region in Fig.~\ref{negative-region}, which corresponds to
positivity of the quantity $\Delta$ and therefore to fulfilment of
the condition \eqref{eq:48}, gets larger with growing time. In
Fig.~\ref{positive-region} the quantity $\Delta$ is plotted as a
function of $\tau$ and of the ratio $r_{+}/r_{-}$. For the case of
the spin-boson model considered in \cite{Maniscalco2007a}, where
the decay constants $\kappa_{\pm}$ are expressed in terms of the
mean number of excitations of the reservoir at the frequency
$\omega$ of the two-level system, taking Eq.~\eqref{eq:60} into
account the constraint Eq.~\eqref{eq:59} is equivalent to
\begin{equation}
   \label{eq:61}
   \beta \hbar \omega \leq \ln (2+\sqrt{3}).
\end{equation}
The time evolution is therefore completely positive only for
reservoir temperatures above a certain threshold given by
$k_{B}T\approx 0.8 \hbar \omega$.

In particular for $r_{+} = r_{-}$, which corresponds to an
infinite temperature reservoir, complete positivity is granted on
the basis of the classical condition Eq.~\eqref{eq:51} only, as
already discussed in Sec.~\ref{sec:n-level-system}, and confirmed
by other approaches \cite{Budini2004a,Maniscalco2007a}. In fact in
this case $T_{++}(t)=T_{--}(t)=T(t)$ and
$g_{++}(t)=g_{--}(t)=g(t)$, so that the condition \eqref{eq:48}
reduces to $T(t)\geq g(t)$, which in our approach is automatically
known to be true because of the probabilistic interpretation of
the quantities involved, and corresponds to the inequality (22) in
Ref.~\cite{Maniscalco2007a}.

\section{Conclusions}\label{sec:conclu}
We have constructed a quantum mechanical generalization of the
classical concept of semi-Markov processes and discussed the
basic features of the resulting quantum semi-Markov processes,
including, in particular, the formulation of mathematical criteria
for the complete positivity of the corresponding quantum dynamical
maps. The main motivation of our study was the development of a
structural characterization of non-Markovian dynamics for a large
class of quantum processes that is relevant for physical
applications. The approach followed here could indeed be
particularly useful in applications for which a microscopic
system-environment approach is technically too complicated or
impossible, guiding the phenomenological construction of the
memory kernel.

It it important to stress that the class of quantum semi-Markov
processes considered here do contain the Markovian limit as a
straightforward special case as shown at the end of
Sec.~\ref{sec:CP-conditions}, thus providing a natural
generalization of quantum Markov processes. In our derivation of
the various conditions for complete positivity we have made some
specific assumptions in order to obtain explicit constraints. For
example, we have assumed a certain structure for the Hamiltonian
part and for the loss term of the memory kernel.  More general
quantum semi-Markov processes can be considered, and will be the
object of future investigations. In particular, we notice that the
conditions obtained in Sec.~\ref{sec:CP-conditions}, which are
only sufficient, could be too stringent in certain physical
applications. It is therefore important to study further examples
in order to decide whether or not a generalization of these
criteria is necessary in practice. This is particularly true for
the case of temporarily negative memory functions (see
Sec.~\ref{sec:NEGATIVE-MEMORY}). However, as is shown in the
examples of Sec.~\ref{sec:examples}, for specific cases one can
formulate conditions for the complete positivity which are not
only sufficient but also necessary, thus leading to a complete
characterization of physically admissible memory kernels. This has
been done for the memory kernel of Eq.~\eqref{eq:41} which
describes a two-level system interacting with a bosonic reservoir,
extending the partial analysis given in
\cite{Budini2004a,Maniscalco2007a}. It has been shown that
Eq.~\eqref{eq:59}, together with the constraint \eqref{eq:51} for
the allowed region of decay constants, is indeed a necessary
condition for complete positivity, and numerical evidence strongly
suggests that this condition is also sufficient. This criterion
for complete positivity can also be understood as a bound on the
temperature range over which the model can give physically
well-defined results.

\begin{acknowledgments}
One of us (HPB) gratefully acknowledges financial support from the
Hanse-Wissenschaftskolleg, Delmenhorst.
\end{acknowledgments}

\end{document}